\documentclass[sigconf]{acmart}
\usepackage{amsmath}
\usepackage{amssymb}
\usepackage{graphicx}
\usepackage{setspace}
\usepackage{xcolor}
\usepackage{subfigure}
\usepackage{courier}
\usepackage{enumitem}
\usepackage[font=normal,skip=0pt]{caption}
\usepackage{xspace}
\usepackage{booktabs} 
\setcopyright{none}

\acmConference[CIDR'21]{}{}{Chaminade CA, USA}
\DeclareRobustCommand{\system}{\mbox{\sc Leam}\xspace}
\DeclareRobustCommand{\vta}{\mbox{\sc VTA}\xspace}
\DeclareRobustCommand{\vita}{\mbox{\sc VITA}\xspace}

\DeclareRobustCommand{\vitaframe}{\mbox{\sc VITAframe}\xspace}

\newcommand{\stitle}[1]{\vspace{0.2em}\noindent\textbf{#1}}

\newcommand{\emtitle}[1]{\vspace{0.15em}\noindent\underline{\em #1}}
\newcommand{\hide}[1]{}

\newcommand{\candidate}[1]{{#1}}
\newcommand{\papertext}[1]{}%\textcolor{blue}{#1}}

\definecolor{corange}{HTML}{C31D00}

\newcommand{\eg}{{\itshape e.g.}, }
\newcommand{\ie}{{\itshape i.e.}, }

\newcommand{\squishlist}{
  \begin{list}{$\bullet$}
           { \setlength{\itemsep}{0pt}
             \setlength{\parsep}{2pt}
             \setlength{\topsep}{2pt}
             \setlength{\partopsep}{0pt}
           }
}
\newcommand{\squishend}{\end{list}}

\newcommand{\squishenum}{
       \begin{enumerate}
               { \setlength{\itemsep}{-50pt}
                       \setlength{\parsep}{0pt}
                       \setlength{\topsep}{0pt}
                       \setlength{\partopsep}{0pt}
               }
       }
\newcommand{\squishenumend}{\end{enumerate}}

\newcommand{\code}[1]{\texttt{\small #1}}

\begin{document}
\title{{\system}: An Interactive System for In-situ Visual Text Analysis}

\settopmatter{printacmref=false, printfolios=true,printccs=false
}

\renewcommand\footnotetextcopyrightpermission[1]{}

\author{Sajjadur Rahman}
\affiliation{%
  \institution{Megagon Labs}
}
\email{}

\author{Peter Griggs}
\affiliation{
  \institution{Megagon Labs \& MIT}
}
\email{}

\author{\c{C}a\u{g}atay Demiralp}
\affiliation{
  \institution{Megagon Labs}
}
\email{}

\renewcommand{\shortauthors}{}

\begin{abstract}
With the increase in scale and availability of digital text generated on the web, enterprises such as online retailers and aggregators often use text analytics to mine and analyze the data to improve their services and products alike.  
Text data analysis is an iterative, non-linear process with diverse workflows spanning multiple stages, from data cleaning to visualization. Existing text analytics systems usually accommodate a subset of these stages and often fail to address challenges related to data heterogeneity, provenance, workflow reusability and reproducibility, and compatibility with established practices.  Based on a set of design considerations we derive from these challenges, we propose \system, a system that treats the text analysis process as a single continuum by combining advantages of computational notebooks, spreadsheets, and visualization tools. \system features an interactive user interface for running text analysis workflows, a new data model for managing multiple atomic and composite data types, and an expressive algebra that captures diverse sets of operations representing various stages of text analysis and enables coordination among different components of the system,  including data, code, and visualizations. We report our current progress in \system development while demonstrating its usefulness with usage examples. Finally, we outline a number of enhancements to \system and identify several research directions for developing an interactive visual text analysis system.
\end{abstract}

%\keywords{Interactive text analysis, text data management, text algebra.}

\maketitle

\pagestyle{plain}

\section{Introduction}\label{sec:intro}
With the rapid  growth of the e-commerce economy, the internet has become the platform for many of our everyday activities, from shopping to dating, to job searching, to travel booking.  A recent study projects the worldwide e-commerce sales to be around six trillion dollars by 2023~\cite{ecommerce}, nearly a 50\% increase over the current market. This growth has contributed to the proliferation of digital text, particularly user-generated text (reviews, Q\&As, discussions), which often contain useful information for improving the services and products on the web. Enterprises increasingly adopt text mining technologies to extract, analyze, and summarize such information from the unstructured text data.
Like data analysis at large, text data analysis also benefits from interactive workflows and visualizations as they facilitate accessible, rapid iterative analysis. Therefore, we characterize the text data analysis process more formally as visual interactive text analysis (\vita hereafter). A challenging aspect of \vita is its iterative and non-linear nature---it is a multistage process that involves tasks like data preprocessing and transformation, model building, hypothesis testing,
and insight exploration, all of which require multiple iterations to obtain satisfactory outcomes.

While there are many commercial and open-source tools that support various stages of \vita~\cite{liu2018bridging, mlbazaar}, none of these capture the end-to-end \vita pipeline. 
For example, spreadsheets are suitable for directly processing and manipulating  data, computational notebooks enable flexible exploratory analysis and modeling, and visualization systems, typically based on chart templates, facilitate quick interactive visual analysis. There are also many customized text analytics tools~\cite{liu2018bridging}, often in the form of research prototypes, that support specific use-cases like review exploration~\cite{zhang2020teddy}, sentiment analysis~\cite{kucher2018state}, and text summarization~\cite{carenini2006interactive}.
Unfortunately, none of these solutions accommodate the inherently cyclic, trial-and-error-based nature of \vita pipelines in an integrated manner.

% \begin{figure}[!htb] 
\begin{figure}[tbp] 
% \vspace{-10pt}
\vspace{-5pt}
  \centering
  \includegraphics[width=\linewidth]{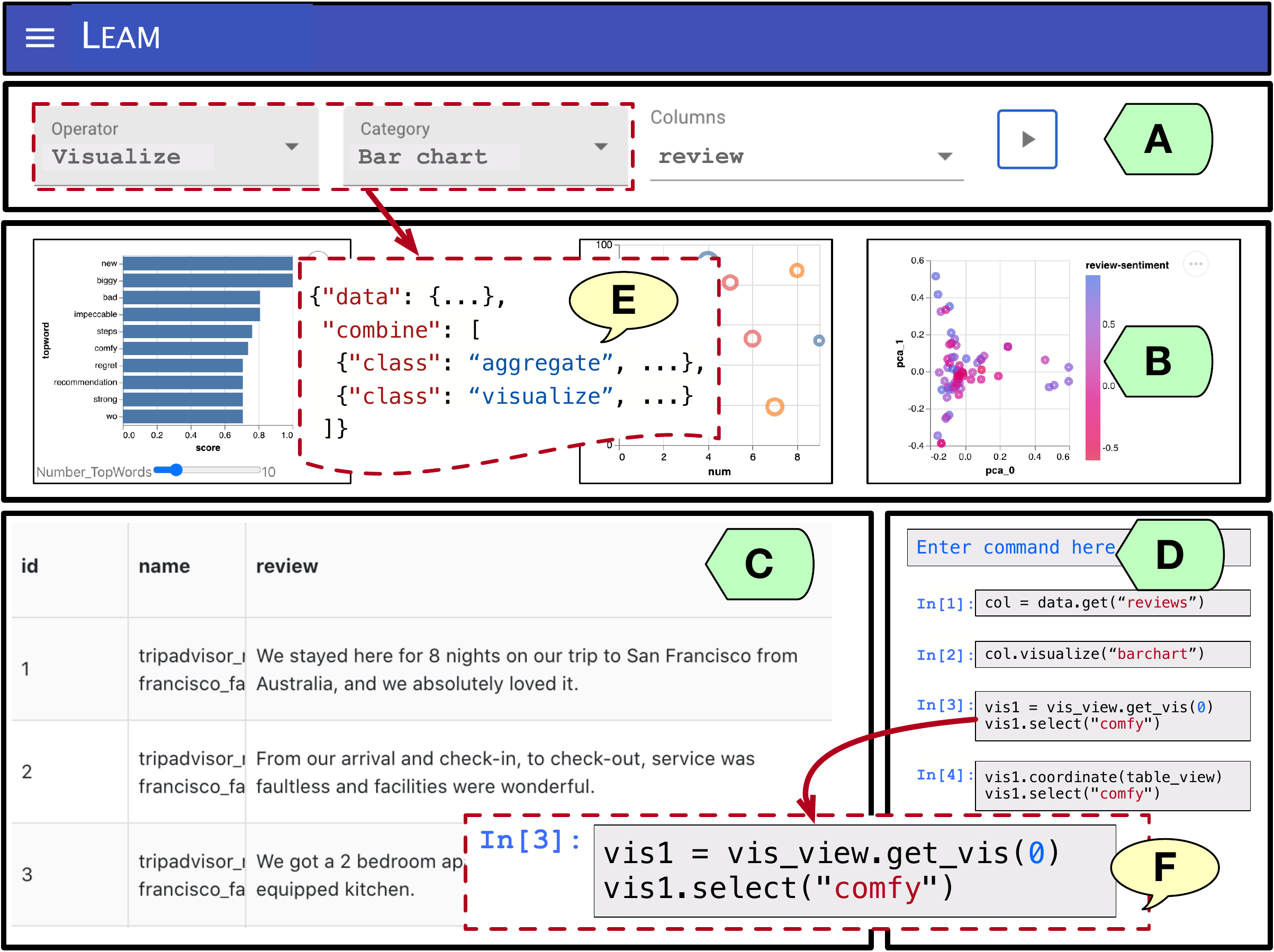}
  \caption{\small \system user interface. (A) Operator View enables users to perform visual text analytics (\vita) operations using drop-down menus, (B) Visualization View 
  holds a carousel of interactive visualizations created by users, (C) Table View displays the data and its subsequent transformations, and (D) Notebook View allows users to compose and run \vita operations using a declarative language. Inset (E) shows the \vta \emph{json} specification for the bar chart operator in Operator View. Inset (F) shows a declarative \vta command for interacting with the bar chart from Notebook View. \label{fig:fe}} 
 \vspace{-15pt}
\end{figure}

While supporting the entire \vita life-cycle within a single system does seem natural, developing it leads to several challenges related to (a) extensibility and expressivity of \vita workflows, (b) their continuity and reproducibility, (c) data heterogeneity and provenance, and (d) synchronization of user interactions. We document these challenges in Section~\ref{sec:example}. These challenges are a culmination of (1) our conversations with practitioners while working in an industry research lab, part of a larger company with more than three hundred e-commerce subsidiaries, (2) prior research, particularly those  reporting from interview studies on data analysis workflows~\cite{zhang2020teddy,lee2020demystifying}, and (3) our experience in developing and evaluating interactive data systems. As such, the list of challenges here is intended to be a useful guide informing research and development on text analytics systems, not a comprehensive enumeration, and inevitably reflects our personal taste. We propose a set of design criteria (desiderata) to address the challenges of a \vita system (Section~\ref{sec:example}). One crucial theme underlying these criteria is the consideration of data analysis as a single continuum, not as discrete steps of tasks performed in isolation.

In this paper, we introduce \system, a one-stop-shop for visual interactive text analysis. \system combines the advantages of spreadsheets, computational notebooks, and visualization tools by integrating a Notebook View with interactive views of raw and transformed data (Figure~\ref{fig:fe}). A key component in the design of \system is a visual text algebra (\vta) that enables users to specify complex \vita operations over heterogeneous data and their visual representations; either using a declarative language in the Notebook View (Figure~\ref{fig:fe}F) or by creating operators in the front-end that translates to \emph{json}-style \vta specifications (Figure~\ref{fig:fe}E). Through usage examples, we demonstrate the expressivity of \vta and how it enables \system to support diverse tasks ranging from data cleaning to visualization. Moreover, to facilitate efficient execution of \vta on heterogeneous data, we introduce a new data model extending dataframes called \vitaframe. Based on our experience in developing \system,
we have identified several system design challenges related to storage model efficiency, scalable computation of \vita workflows, data and workflow versioning, and workflow optimization. To address these challenges, we identify several research directions that may enhance the capabilities of \system. We have made the current version of \system open-source at~\url{https://github.com/megagonlabs/leam}.

The goal of this paper is to explore challenges in and design considerations for \vita systems development,
present vital components of \system that enable interactive text analysis (\eg \vitaframe and \vta), and identify
some concrete research directions based on our experience of developing \system.

% recommendations 

\begin{figure}[tbp] 
%  \vspace{-10pt}
  \centering
  \includegraphics[width=\linewidth]{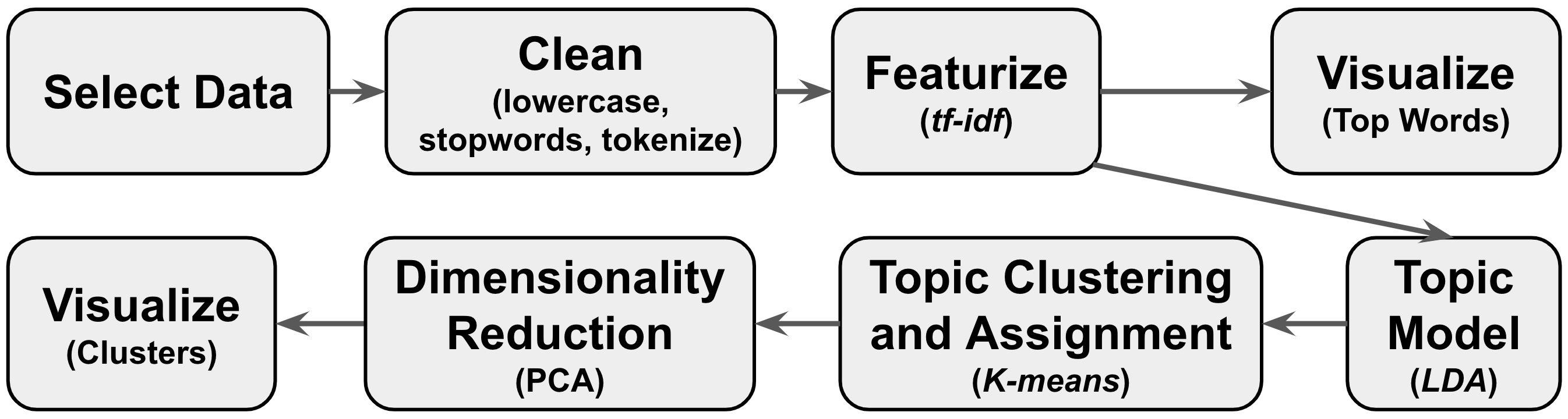}
  \caption{\small An example \vita use-case: topic exploration.\label{fig:workflow}} 
  \vspace{-20pt}
\end{figure}

\candidate{\section{\vita Challenges}}
\papertext{\section{\vita: Challenges and design goal}}
\label{sec:example}
\stitle{Motivating Example.} 
Ada, a data scientist in the e-commerce department of a retail business, has been tasked to analyze customer reviews of products purchased from their website. Ada would like to capture the underlying topics by performing topic modeling and clustering to characterize the review corpus better. Figure~\ref{fig:workflow} captures the use-case which involves---(a) preprocessing the data (\emph{clean}), creating feature vectors from the text reviews (\emph{featurize}), creating topic vectors from the corpus (\emph{topic modeling}), clustering reviews into topics (\emph{cluster assignment}), and finally, visualizing the clusters by projecting the topics vectors to lower dimensions (2D) using feature transformation techniques such as PCA (\emph{visualize}). In practice, the workflow may be non-linear, and each step may require multiple passes and different tools. In the process, visualizations are useful not only for exploratory analysis or final presentation but also for every other step---\vita workflows resemble the \emph{read-eval-print loop} (REPL) approach where users perform incremental operations on data and examine intermediate results.We now characterize the challenges of the existing \vita workflows in the context of this use-case as follows:

\emtitle{C1. Workflow discontinuity.} As mentioned in Section~\ref{sec:intro}, data scientists lack tools 
that support different \vita operations and workflows within an integrated environment.
For example, to define the data cleaning rules, Ada first visually inspects the data using tools like spreadsheets. Next, they execute those rules in a computational notebook, \eg Jupyter. Upon inspecting the data in the spreadsheet, Ada may revise the rules in the notebook. To visualize top-ranked words after the featurization step (\eg a bar chart of words ranked by their TF-IDF scores), they need to either use dedicated visualization tools or write scripts in the notebook. Therefore, even completing simple tasks may require accessing different tools, which can be cumbersome user experience due to, for example, the logistical and cognitive overhead of context switching.

\emtitle{C2. Limited coordination.} \vita necessitates coordination among different views (\eg between visualization and raw data).
The high dimensional text data can be difficult to interpret and users
often map different facets of the data to visualizations for better interpretability. 
However, without coordination between perceptual components and the data space, understanding the relations between the facets of the same entities on demand can be challenging. 
For example, say Ada wants to inspect which reviews contain a top-word in the bar chart (generated after featurization). However, visualizations in notebooks or visualization tools are decoupled from the data. As a result, Ada has to either open and then filter the data in a spreadsheet, or programmatically filter the data from the notebook to inspect the relevant reviews.
Therefore, the lack of coordination impacts both workflow continuity and the user's ability to explore data effectively.

\emtitle{C3. Data types and workflow diversity.} \vita workflows deal with heterogeneous data (\eg text, visualizations) and workflows (\eg in use-cases like text summarization, sentiment analysis). While there are a number of \vita tools for specific workflows~\cite{liu2018bridging}, more often than not these tools use a stack of independent solutions for data storage and
processing glued together by scripting languages like Python and R.
These bespoke solutions may not capture all \vita requirements like direct data manipulation and interactive visual coordination. As a result, users are often forced to develop new and heavily customized systems on top of these solutions.

\emtitle{C4. On demand workflow authoring.} 
\vita workflows contain a variety of operations, \eg cleaning, featurization, interactive visualization, classification. Similar to relational~\cite{codd} or data visualization algebra~\cite{satyanarayan2016vega}, \vita operations with similar objectives can be grouped into high-level categories. Moreover, operations in different categories can be combined to compose new operation pipelines. For example, cleaning and featurization can be combined into a preprocessing pipeline. As existing systems lack any formalization of the operations and their application, the onus is on the user to design the optimal analytics workflow for different use-cases.

\emtitle{C5. \vita Session management.} As demonstrated in the usage example, \vita workflows inspire the trial-error-correct-style iterative approach---users often need to reproduce previous steps of the workflow, make updates, and rerun the subsequent steps. 
Therefore, ensuring reproducibility of 
\vita sessions require
management of dataset versions produced by various operations, the operation logs, and different states of
and interactions on the visual representations of the data. 
Prior work from the data management community focused on versioning structured datasets~\cite{huang2017orpheusdb}, versioning code for debugging workflows~\cite{brachmannbyour,miao2016provdb} and managing deep learning models~\cite{miao2016modelhub}.
However, these systems lack support for versioning an end-to-end \vita workflow involving heterogeneous data types and user interactions spanning multiple views.

\stitle{Design goals.} We propose the following design principles to address the challenges related to \vita:

\emtitle{D1. In-situ analytics.}
\vita systems should provide a one-stop-shop (\textbf{C1}) where users can directly
manipulate (spreadsheets) and visualize (visualization tools) data while writing scripts (notebooks) to immediately view the effects on data and visualizations without context switching between tools.

\emtitle{D2. Multi-view coordination.}
Beyond integrating multiple views within a single interface, \vita systems should enable coordination between these views (\textbf{C2}).
Multiple coordinated views capture the context of the user's exploration across different views~\cite{wang2000guidelines} and help users understand the data better as they view it through different connected representations.

\emtitle{D3. Heterogeneous data management.}
\vita systems should support heterogeneous data types (\eg texts, visualization), treating them as first-class citizens of the underlying data model (\textbf{C3}). Instead of developing bespoke data management solutions, \vita systems should adapt their underlying storage model to
accommodate these data types and also enable a tight coupling between the data model and the analytical workflows to ensure fast and efficient data access.

\emtitle{D4. Expressivity and accessibility.}
\vita systems should provide an expressive specification language to represent and communicate the entire breadth of workflows within the domain (\textbf{C4}). 
%The specification language should characterize the data domain with heterogeneous data types, abstract operations into high level categories, define rules for synthesizing new operations, and capture the coordinated interactions between multiple views. 
Moreover, the specification language should be accessible to existing tools to allow more expressive operations. 
For example, the specification language can be packaged as a Python library with an interactive widget with support for a subset of \vita operations in a computational notebook.

\emtitle{D5. Provenance.}
\vita systems should support advanced provenance tracking for heterogeneous data types and various workflows to ensure reproducibility and encourage workflow and data re-use. 
Moreover, these systems should track user interactions on visual components to enable versioning of states of and dependencies among different views.

\section{\system: In-situ Analysis}\label{sec:system}
\begin{figure}[tbp] 
%\vspace{-20pt}
  \centering
  \includegraphics[width=\linewidth,trim={0 15 0 10},clip]{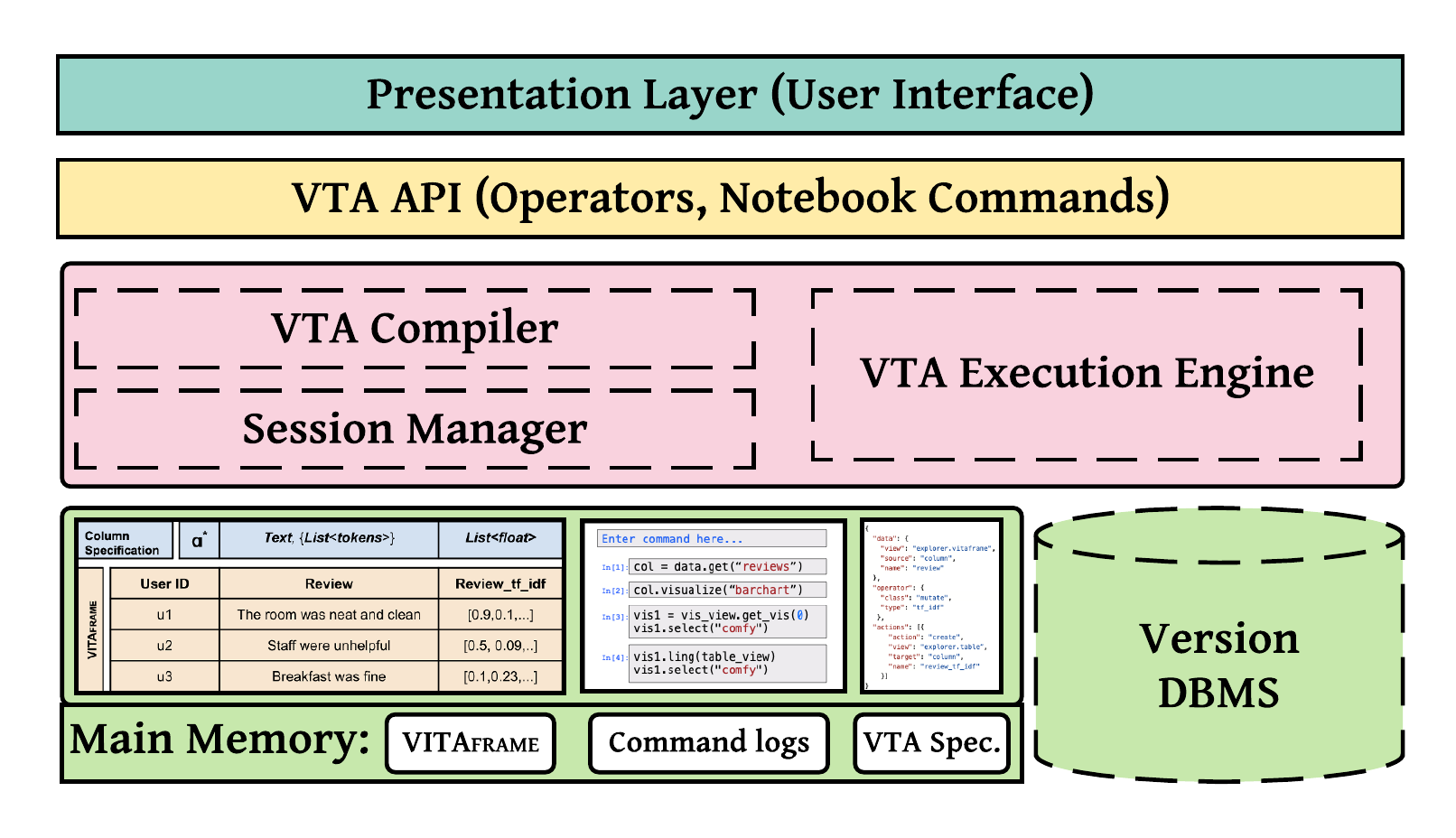}
  \caption{\small \system system architecture.\label{fig:arch}}
 \vspace{-15pt}
\end{figure}

\system backend features an in-memory dataframe, a versioning database, a compiler for translating \vta commands that the execution engine runs, and a session manager to manage data, code, and visualizations. Figure~\ref{fig:arch} shows the overview of the \system system architecture. The components with dashed borders (``- -'') are partially implemented and require further refinement. We discuss the front-end components later.

\subsection{Data Model}
The underlying data structure of \system is a dataframe. A dataframe is a multidimensional array, $A_{mn}$, with a vector of row labels, $R = \{R_1, R_2, \ldots, R_m\}$ and a vector of column labels, $C = \{C_1,C_2,\ldots,C_n\}$~\cite{modin}. 
We opted for dataframes instead of a database since 
\vita tasks
(\eg featurization, classification) cannot always be conveniently performed inside the database~\cite{lajus2014efficient}.
Dataframes are widely used in
exploratory data analysis, including \vita due to their coverage of a wide variety of data analysis operators~\cite{modin}. 
\candidate{The \emph{Pandas} dataframe API within Python (pandas.pydata.org)
has been downloaded more than 300 million times and served as a
dependency for over 265,000 repositories in GitHub.} 
They also do not require data to be defined schema-first allowing flexibility in supported data types and data structures. Finally, dataframes provide a functional interface suitable for REPL-style \vita workflows to perform ``quick and dirty'' analysis.

\begin{figure}[!htb] 
\vspace{-10pt}
  \centering
  \includegraphics[width=\linewidth]{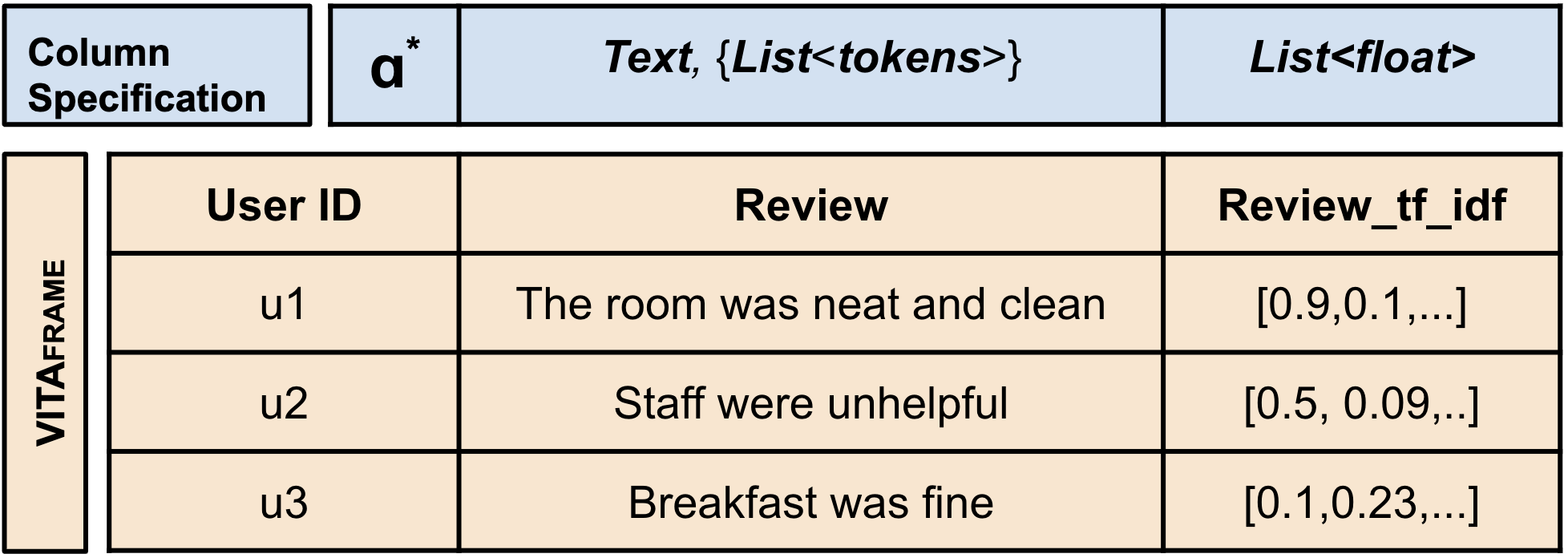}
  \caption{\small \vitaframe with column and metadata schema specification. The Review column is of type $\mathbf{Text}$ and the metadata, the collection of unique tokens, is of type $\mathbf{List}(\mathbf{token})$.\label{fig:column_spec}} 
  \vspace{-13pt}
\end{figure}

\begin{figure*}[tbp]
\centering
\small
\begin{tabular}{ c c c}
  \includegraphics[width=0.3\linewidth,height=0.18\linewidth]{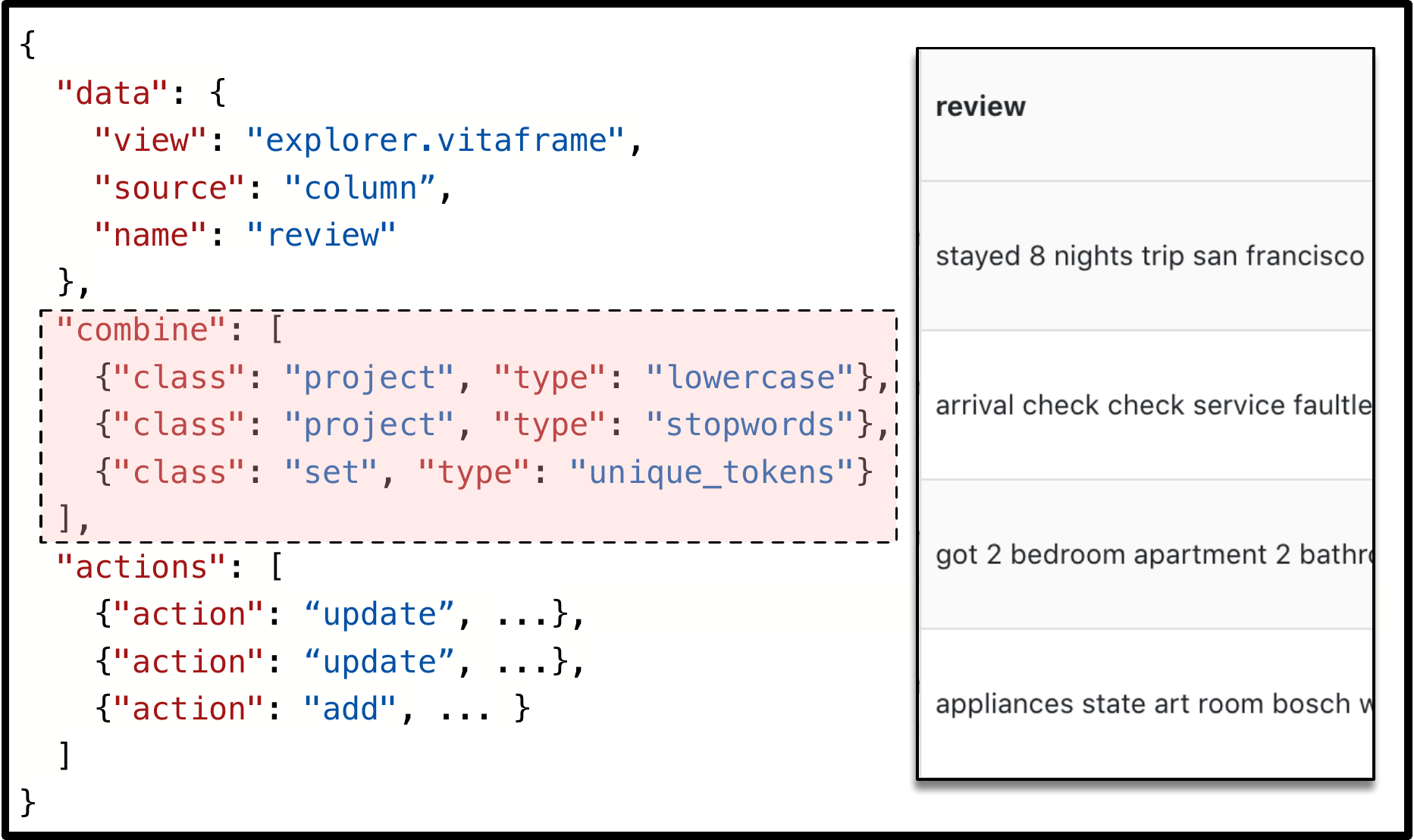}   & 
  \includegraphics[width=0.3\linewidth,height=0.18\linewidth]{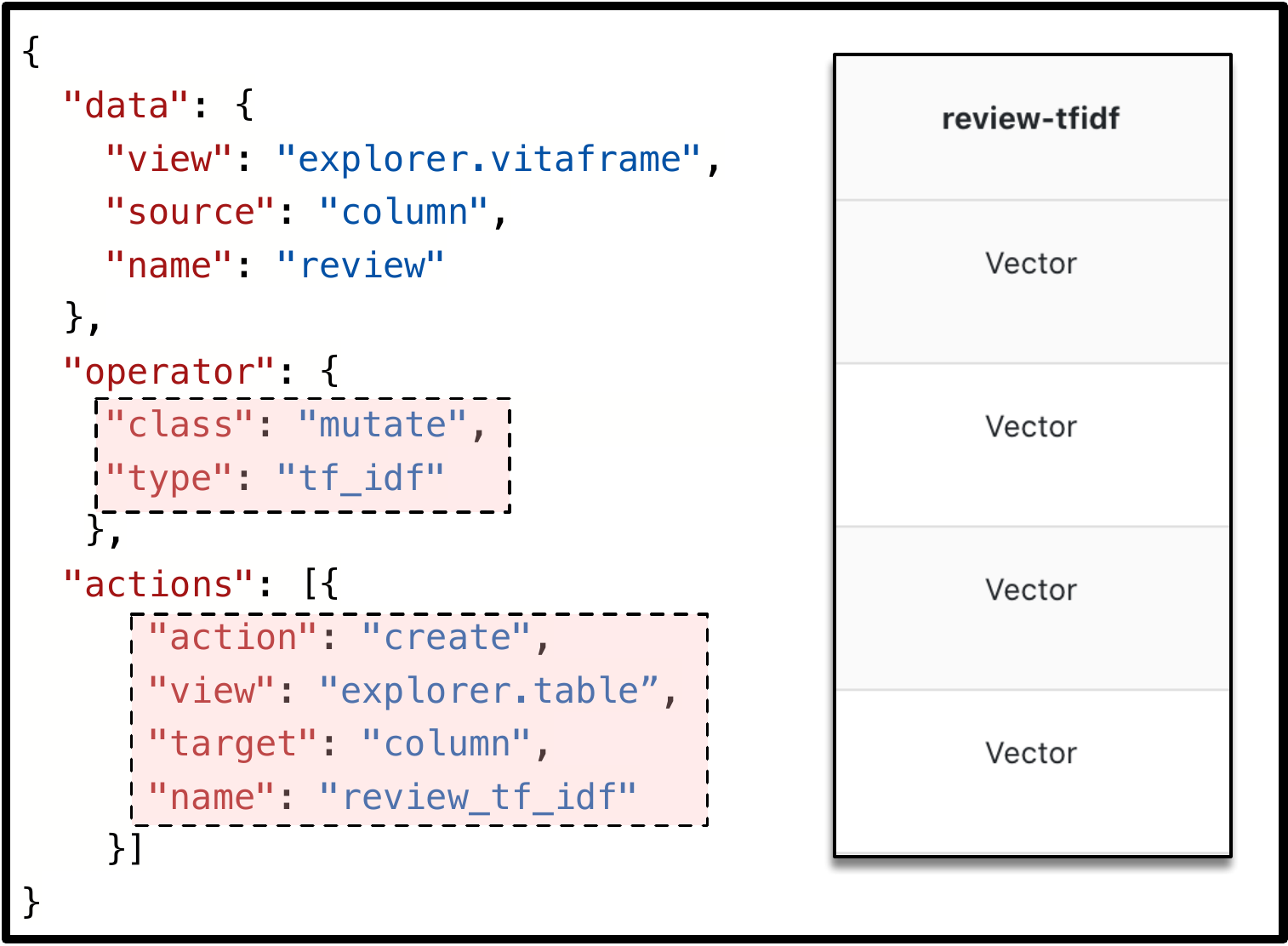} & 
  \includegraphics[width=0.4\linewidth,height=0.18\linewidth]{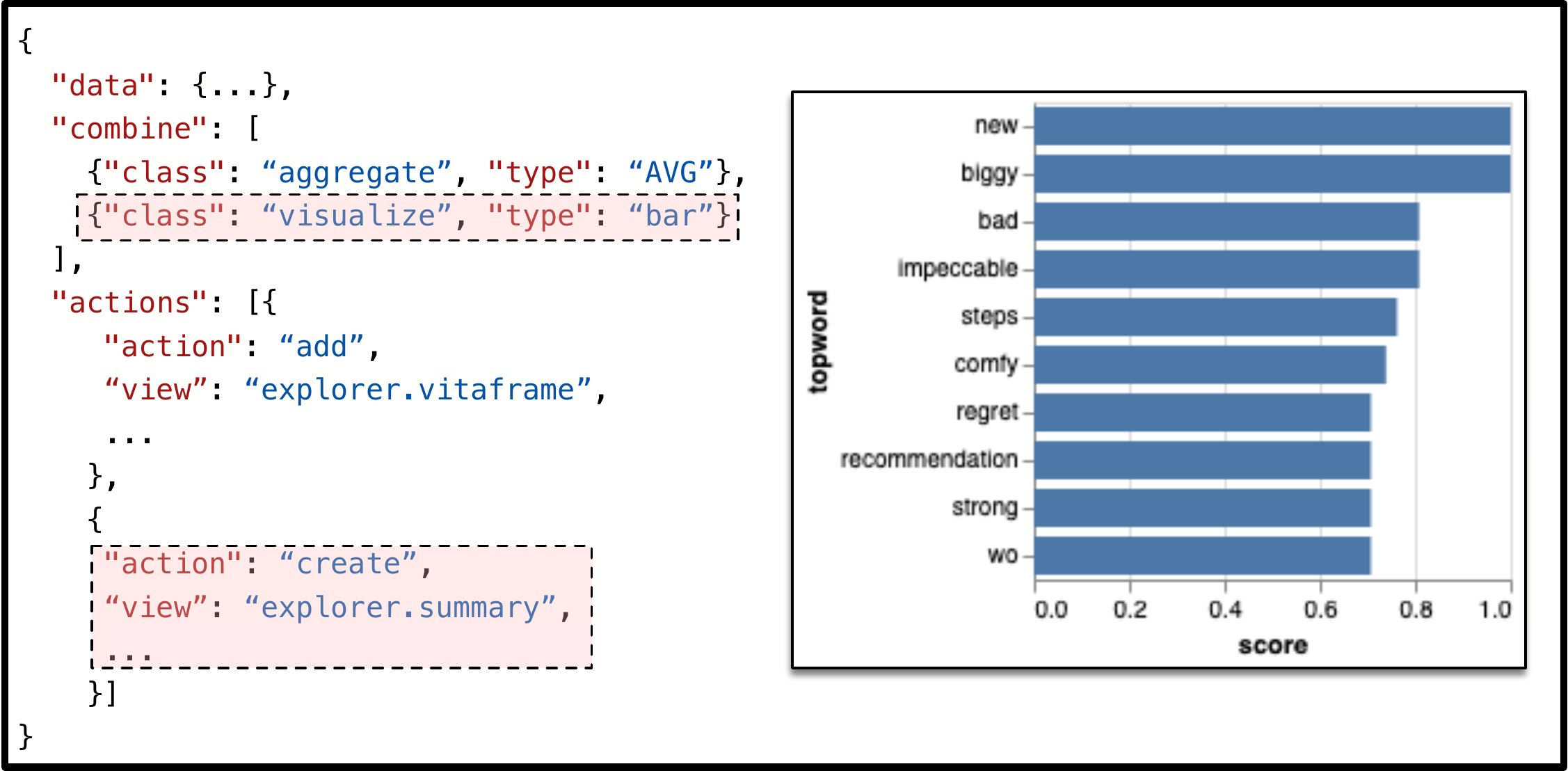} \\
     (a) Clean and generate metadata & 
     (b) Create TF-IDF feature vector &
     (c) Visualize top-words by TF-IDF score
\end{tabular}
    \caption{\small \vta specification. (a) Create a composite cleaning operator ( \code{lowercase} and \code{remove\_stopwords}) using \code{combine} and generate metadata (\code{unique\_tokens}). (b) Create TF-IDF vectors from reviews using \code{mutate} operator (added as a new column in Table View). (c) Create a bar chart by combining \code{aggregate} (average TF-IDF score for each token) and \code{visualize} operator.}
\papertext{\vspace{-15pt}}
    \label{fig:use-case-a}
\end{figure*}

To support \vita use-cases,  we assign a schema to dataframe columns---each column $C_i \in C$ may have a schema defined over a set of data domains, $D = \{d_1, d_2, \ldots\}$ that spans heterogeneous data types like text, visualizations (\textbf{D3}). We call this data structure a \vitaframe. We discuss the underlying data domain for \vita in Section~\ref{sec:vta}.
The REPL-style \vita workflows involve users creating and examining intermediate results that are also from the data domain, $D$. An intermediate result with one-to-one correspondence with a \vitaframe column
(\eg $n$ reviews are featurized into $n$ feature vectors) is added as a new column. However, these results may not have one-to-one correspondence (\eg dictionary of words in the set of $n$ reviews) and are stored in a separate data structure as metadata of the corresponding column.
Therefore, for each column $C_i \in C$ in a \vitaframe, there is a schema specification function that assigns a domain $d_j \in D$ to the column and a domain $d_k \in D$ for each column metadata (see Figure~\ref{fig:column_spec}).
\candidate{For example, in Figure~\ref{fig:column_spec}, the column \emph{Review}'s type is $Text$ and metadata type is $\mathbf{List}(\mathbf{token})$ (\ie the list of unique tokens in the text reviews).} 
Metadata are often used for computing aggregate statistics and visualizations.
\candidate{For example, to visualize TF-IDF score of top ranked words in the reviews, a user can first construct a new metadata of type $\mathbf{Dictionary}(\mathbf{token}, \mathbf{float})$ that contains the average TF-IDF score of the unique tokens. Then they can use the metadata to create and visualize a ranked list of words as a bar chart (see Figure~\ref{fig:use-case-a}c).
Note that there can be columns with no metadata.}

\subsection{Computation Model}\label{sec:computation}
\stitle{\vta Compiler.} 
In Section~\ref{sec:vta}, we present \vta,  an algebra for specifying \vita operations. \system compiles the user interactions on Operator View (see in Figure~\ref{fig:fe}a) or \vta commands in Notebook View (see in Figure~\ref{fig:fe}d) into \vta specifications. However, the \vta specifications are incomplete, in the sense that they may omit details ranging from visual encoding such as fonts, line widths to input data type.
To resolve these ambiguities \system currently uses a rule-based compiler that translates 
a \vta specification into lower-level operators for the execution engine to run backend computation (see Section~\ref{sec:vta}).

\stitle{Execution Engine.} 
The \vita execution engine takes the following input generated by the \vta compiler: (1) input data schema in $D$ 
and (2) translated \vta specifications.
\papertext{The text analysis operations---data preprocessing,
featurization, 
feature transformation/selection),
estimation),
and more advanced post-processing operations
like anomaly detection---are mapped to existing ML and NLP libraries like Spacy, Scikit-learn. While other \vta operators related to visual coordination are mapped to built-in implementations (see Section~\ref{sec:vta}).}
\candidate{The text analysis operations---data preprocessing (\eg cleaning),
feature extraction (\eg TF-IDF featurization), 
feature transformation/selection (\eg PCA),
estimation (\eg classify),
and more advanced post-processing operations
(\eg anomaly detection)---are mapped to existing ML and NLP libraries like Spacy, Scikit-learn. While other \vta operators related to visual coordination are mapped to built-in implementations.}

\stitle{Version Control.}
After each operation, \system checkpoints the current state of the visualizations, \vitaframe, and notebook commands. The development of a fully functional system for fine-grained (\ie at operation level) versioning is currently in progress (\textbf{D5}). 
 We discuss how to support fine-grained version management of a \vita session involving heterogeneous data types, notebook commands, and user interactions in
 Section~\ref{sec:discussion}.

\subsection{\system Front End}\label{sec:ui}
\system user interface has four components---Operator View, Table View, Visualization View, and Notebook View (see Figure~\ref{fig:fe})---which enables users to perform in-place text analytics (\textbf{D1}). Users can perform various \vita operations using the operators in Operator View (see Figure~\ref{fig:fe}a). For example, cleaning the data in Table View, adding visual summaries in Visualization View. Table View (see Figure~\ref{fig:fe}c) design is inspired by traditional spreadsheets and tabular data visualization tools~\cite{spenke1996focus} and enables users to directly operate on the data.
Table View data can be transformed into visual summaries like bar charts and scatterplots using the visualization operators or \vta commands (see Figure~\ref{fig:fe}). 
\candidate{Moreover, a cell in Table View can also be a visualization, similar to tabular data analytics tools~\cite{spenke1996focus}.} 
The visualizations in Visualization View (see Figure~\ref{fig:fe}b) are displayed as a carousel of charts. Interactive visualizations are generated by translating the visualization operators or commands to Vega-Lite specifications~\cite{satyanarayan2016vega}.
Notebook View design (see Figure~\ref{fig:fe}d) is inspired by computational notebooks and enables users to author \vita workflows using a Python-based \vta library. The different views in \system can be linked---interactions on one view are reflected in other views (\textbf{D2}). Through \vta, users can declaratively specify interactive coordination (\eg brush-and-linking) between these views that are translated to Vega signals~\cite{satyanarayan2015reactive}. 

\begin{figure*}
\centering
\small
\begin{tabular}{ c c}
  \includegraphics[width=0.5\linewidth,height=0.18\linewidth]{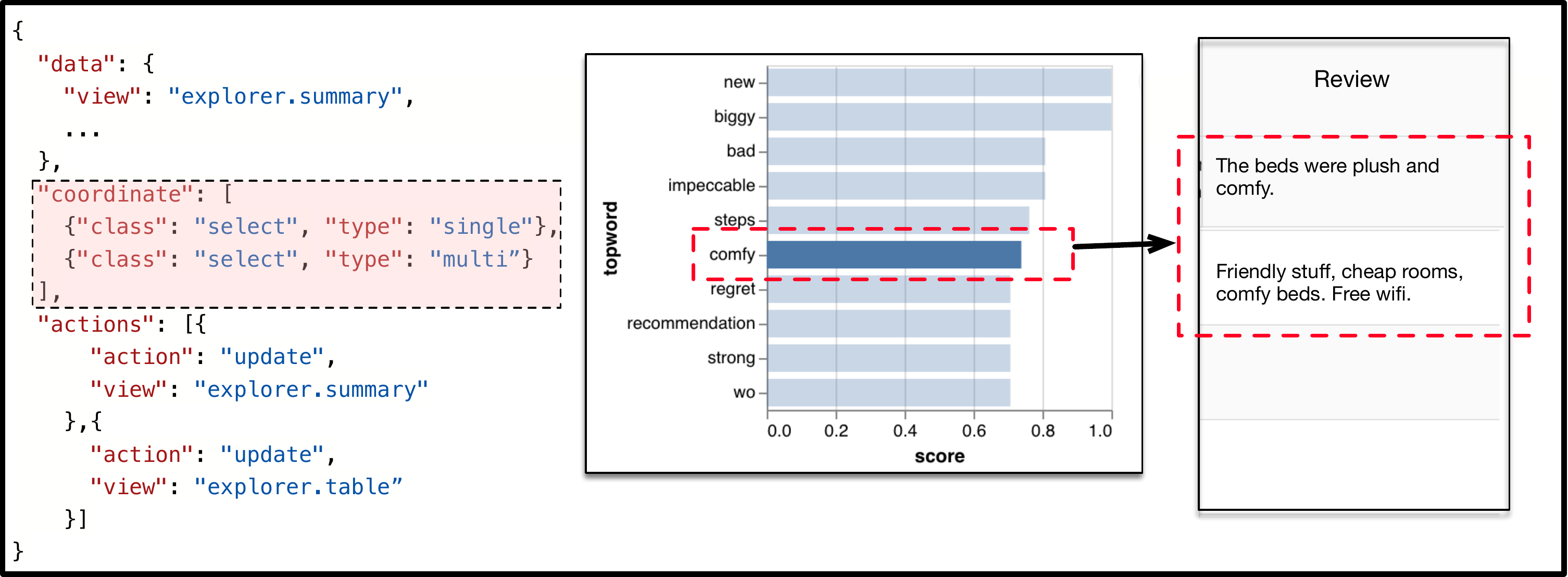}   & 
  \includegraphics[width=0.5\linewidth,height=0.18\linewidth]{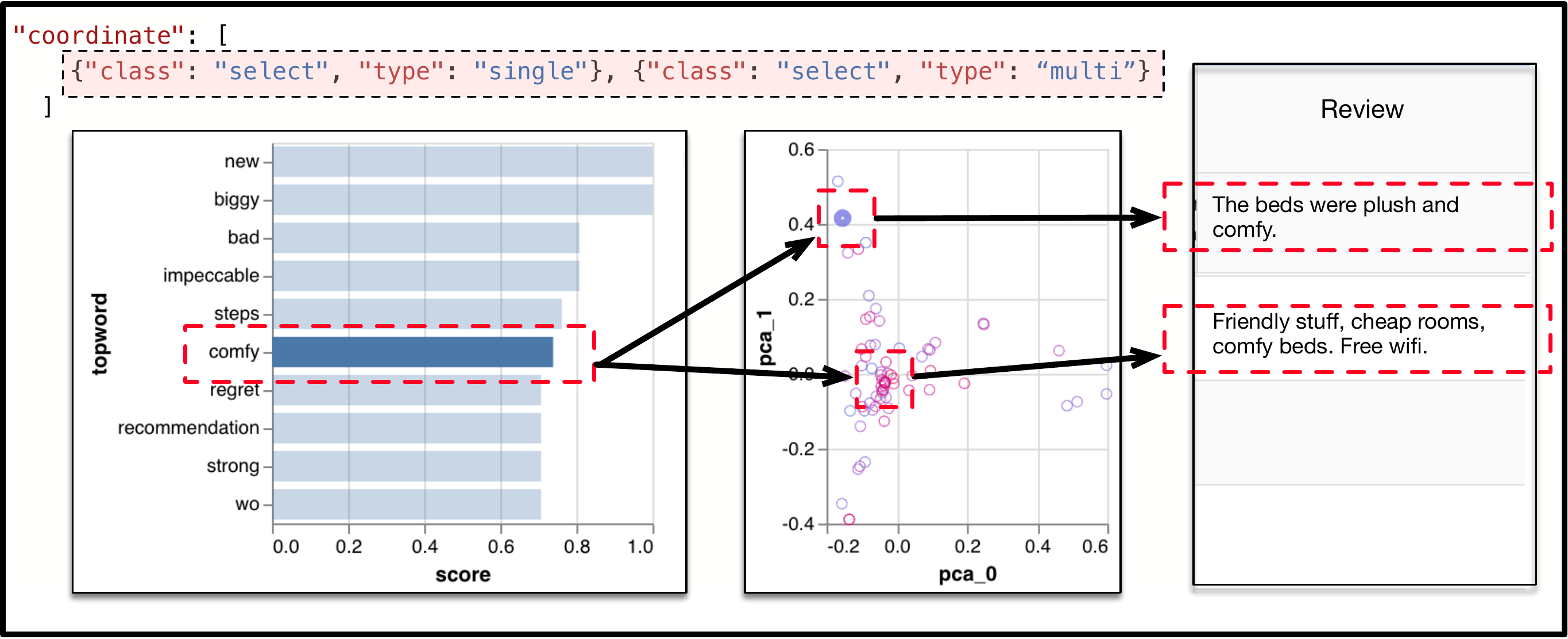} \\
     (a) Coordination of two views & 
     (b) Coordination of three views 
\end{tabular}
    \caption{\small \papertext{Multi-view coordination. (a) Use the \code{coordinate} operator to link the bar chart and Table View. Selecting a bar (word) in the bar chart triggers a \code{filter} by the selected word on Table View . (b) Coordinating the bar chart and scatterplot links the three views. Selecting a bar in the bar chart highlights relevant points in the scatterplot, and filters relevant rows in Table View.}\candidate{Multi-view coordination specification. (a) Use the \code{coordinate} operator to create a unidirectional coordination between the bar chart and Table View. Selecting a bar (word) in the bar chart triggers a \code{filter} operation on Table View by the selected word. (b) Creating a unidirectional coordination between the bar chart and scatterplot automatically links the three views (multi-view coordination). For the previous action on the bar chart, all relevant points in the scatterplot are selected, in addition to filtering Table View.}}
    \label{fig:use-case-b}
    \papertext{\vspace{-15pt}}
\end{figure*}

\section{\vta: An Algebra of \vita}\label{sec:vta}
We now discuss our visual text algebra, \vta, in detail.
We demonstrate how \vta captures various tasks in the usage example in Section~\ref{sec:example}
(see Figure~\ref{fig:fe},~\ref{fig:use-case-a}, and~\ref{fig:use-case-b}). 
%In this section, we first describe \vta's underlying data domain.
%We then define the current operators in \vta that enable users to specify various \vita workflows and interactions within \vita tools such as \system. 

%To the best of our knowledge, an algebra for \vita has never
% been defined previously. 
%However, our work draws inspiration from relational algebra~\cite{codd} and prior work on visualization grammars~\cite{satyanarayan2016vega,satyanarayan2015reactive,stolte2002polaris,bostock2009protovis,bostock2011d3}.

\subsection{\vta Specification}\label{sec:data_domain}
\stitle{Data domain.}
\vita involves many data types, including different forms of text (\eg words, tokens, sentences), complex data types like lists, vectors, and even visualizations. We define the data domain of \vita as $D = \{P, S\}$, where $P$ and $S$ are sets of primitive and synthesized (\ie composite) data types. The domains of primitive data types are taken from, $P = \{\alpha^*, \mathbf{int}, \mathbf{float}, \mathbf{bool}$,
$\mathbf{datetime},\mathbf{List}, \mathbf{Vector}, \mathbf{Dictionary}\}$. The domain $\alpha^*$ is the set of finite strings
over an alphabet $\alpha$. The domains of composite data types are taken from, $S = \{\mathbf{Text}$, $\mathbf{Visualizations}\}$. If a  schema of the data is not specified upfront, the data is initially assumed to be from the domain $\alpha^*$. Each domain $d_i \in D$ includes a generator function $g_i: \alpha^* \rightarrow d_i$ which defines the rule for inferring exact data types of the respective domain. For example, the composite $\mathbf{Text}$ data types (\eg words, tokens, sentences) are generated using a context free grammar~\cite{charniak1997statistical}. The generator function of visualization data types is defined based on Vega-Lite~\cite{satyanarayan2016vega}: $\mathbf{Visualizations} = (data, transforms, mark-type, encodings)$. \vta operators can be specified in a \emph{json} or as declarative commands that are available through a \vta library in Python.

% e now explain how users specify \vita workflows in \vta.
% While explaining the algebra, we also demonstrate how \vta captures various tasks in the usage example in Section~\ref{sec:example}
% (see Figure~\ref{fig:fe},~\ref{fig:use-case-a}, and~\ref{fig:use-case-b}). As shown in these figures, \vta operators can be specified in a \emph{json} format or as declarative commands that are available through a \vta library in Python.

\stitle{Selection operator.}
\vta enables the specification of interaction through selections. 
Selection operations select data points of interest on which subsequent operations in the workflow may be performed (\eg row(s) in Table View, visualization mark(s) in Visualization View). Supported selection types include a single point (\eg a table row, marks like a bar or a circle in a chart),
a list of points (\eg rows, bars, or circles), or an interval of points (\eg ten rows starting from row $i$, circles in scatterplot within $x$-axis range). The selection criteria are specified by a predicate to determine the set of selected points. \code{Filter} is an example of list selection where the selection predicate is the filtering condition. 
\vta also supports similar types of selections on visualizations (Figure~\ref{fig:use-case-b}a). Besides \emph{json} specification, users can also perform such selections by writing commands in Notebook View. For example, Figure~\ref{fig:fe}F shows how a user can select a bar (\eg the word ``comfy'') in the bar chart using a \vta command: 

% \begin{figure}[!htb] 
% \vspace{-10pt}
%   \centering
%   \includegraphics[width=\linewidth]{figures/notebook.pdf}
%   \caption{\small Selecting a bar chart element using \vta command.\label{fig:notebook}}
%  \vspace{-10pt}
% \end{figure}

\stitle{Transformation operators.}
While developing \system, we identified a set of core transformation operators that encompasses the various \vita workflows. 
These transforms manipulate the
components of the selection they are applied to. 
Note that the core operator set is minimal, and there is room for adding more operators to make \vta more expressive (\textbf{D4}). 
The transformation operation has five subclasses: \code{project}, \code{mutate}, \code{aggregate}, \code{set}, and \code{visualize}. The \code{project} operators change the dimensionality (\eg LDA, PCA) or cardinality (\eg stopword removal) of the input data or update the content (\eg lowercase). The \code{mutate} operator generates a new representation of the input data (\eg create a list of tokens or feature vectors from text). The \code{aggregate} operator computes summary statistics of the input data (\eg average review length in a corpus). The \code{set} operations enable set-like operators on the input data (\eg get unique tokens in the corpus). The \code{visualize} operator generates visualizations of data. Figure~\ref{fig:use-case-a} captures the data preprocessing phase of the workflow discussed in Section~\ref{sec:example}. Ada first cleans the reviews using \code{project} operators (Figure~\ref{fig:use-case-a}a),then applies a \code{mutate} operator (\eg TF-IDF feature vector creations) to featurize the reviews (Figure~\ref{fig:use-case-a}b). Finally, Ada computes the average TF-IDF score of each word (\code{aggregate}) and visualizes the top ranking words (\code{visualize}) using a bar chart (Figure~\ref{fig:use-case-a}c). 

\candidate{Each operation takes an input from the given data domain $D$ and generates an output that also belongs to the same domain $D$.
An \emph{action} defines how the resulting output should be used. An action can be of three types: \code{add}, \code{create}, \code{update}. For add action, the output becomes meta-data of the input (\eg metadata of a \vitaframe column). For \code{create}, the output becomes part of the data the user directly operates on (\eg create a new \vitaframe column). For \code{update}, the output essentially replaces the input data (\eg update an existing \vitaframe column).}

\stitle{Composite Transforms.}
\vta currently supports two composite transforms that combine unit transformation operations: \code{combine} and \code{synthesize}. The \code{combine} operator enables users to specify an operation pipeline. In Figure~\ref{fig:use-case-a}a, a user combines two \code{project} operations with a \code{set} operation into a single operation. Similarly, in Figure~\ref{fig:use-case-a}c, a user combines \code{aggregate} and \code{visualize} operators into a single operation for bar chart creation. The \code{synthesize} operator enables users to create new operations from these combinations which can be reused later. For example, a user can \code{synthesize} the previous cleaning pipeline to be a \code{clean} operator which then becomes an operation in the \emph{Operator View} and is used later.

\subsection{Multi-view Coordination}
\papertext{So far, we have explained selections that are defined within a single view. However, selections that involve multiple views cannot be captured by a single-view-based specification. We define a coordination operator called \code{coordinate} that captures multi-view coordination. For example, in Figure~\ref{fig:use-case-b}a, selecting a top-word bar in the bar chart filters relevant reviews in Table View (Visualization View $\rightarrow$ Table View coordination). The \code{coordinate} operator needs to also resolve the mapping between views and composite selections across views.}
\candidate{
So far we have explained selections that are defined within a single view (\eg Table View or Visualization View). However, selections that involve multiple views cannot be captured by the default single-view-based \vta specification. We define a coordination operator called \code{coordinate} that captures such multi-view coordination.
Coordination can be either unidirectional or bidirectional. For example (Figure~\ref{fig:use-case-b}a), selecting a top-word bar in the bar chart and filtering corresponding Table View rows is a unidirectional coordination. However, adding a selection in the reverse direction makes the coordination bidirectional. For example, changing opacity of the bars in the bar chart based on reviews selected in Table View. Currently, \system supports only unidirectional coordination within a single specification. To create a bidirectional specification, users are required to specify two unidirectional specifications. Besides the direction of coordination, the \code{coordinate} operator needs to resolve the mapping between coordinated views and composite selections across views.
}

\papertext{\stitle{Mapping coordination.} 
In Figure~\ref{fig:use-case-b}a, there is a one-to-many mapping from the bar chart to Table View---selecting a bar filters two that reviews contain the word in ``comfy''.
In Figure~\ref{fig:use-case-b}b, there is a one-to-many mapping from the bar chart to scatterplot and a one-to-one mapping from the scatterplot to Table View. Therefore, the \code{coordinate} operator needs to resolve the mapping among views, so that relevant visualization marks are selected/highlighted in respective views. Users can specify the mapping using the \emph{type} tag in respective \code{select} operators of the views. For example, in Figure~\ref{fig:use-case-b}a, the user selects type \emph{single} for the bar chart and \emph{multi} for Table View.}

\candidate{
\stitle{Mapping coordination.} 
In Figure~\ref{fig:use-case-b}a, there is a one-to-many mapping from the bar chart to Table View---selecting a bar may filter multiple reviews that contain the top-word (\eg two reviews contain the word in ``comfy'').
In Figure~\ref{fig:use-case-b}b, there is a one-to-many mapping from the bar chart to scatterplot and a one-to-one mapping from the scatterplot to Table View. Therefore, the \code{coordinate} operator needs to resolve the coordination mapping among views on the fly, so that relevant visualization marks are selected/highlighted in respective views. In our current \vta implementation,
users are required to explicitly specify the mapping using the \emph{type} tag in respective \code{select} operators of the views. For example, in Figure~\ref{fig:use-case-b}a, the user selects type \emph{single} (\ie one) for the bar chart and \emph{multi} (\ie many) for Table View. 
We aim to automate the mapping process in future by resolving the mapping between the underlying data of the respective views.}

\stitle{Resolving composite selections.}  
Users can add multiple single-view selections to a view. For example, in Figure~\ref{fig:use-case-b}b, the user adds a link from the bar chart to the scatterplot, which creates a multi-view coordination among the bar chart, scatterplot, and Table View.
Selecting a bar in the bar chart highlights multiple circles (scatterplot) and reviews (Table View). However, following the top-word selection, the user may select a rectangular area in the scatterplot or select multiple reviews in the table. 
\candidate{It is not clear whether we should deselect the previous selection and only highlight current selection in the scatterplot and update corresponding bar chart and table view selections, \ie always perform independent selection. The other option is to perform a \code{union} or \code{intersection} among the selected data points of all the selections.}
Currently, \system only supports independent selection. We outline advanced composite specifications such as \code{union} and \code{intersection} as future work.

\section{\system Enhancements}
\label{sec:discussion}
% \sajreview{We now outline our vision for \vita framework development.}
We now outline our vision for \system development.

\papertext{\stitle{Efficient storage model.}
While \vitaframe currently supports data that fits in memory, in practice, the datasets will be larger. 
So, the storage layer of \system can enable operations on both in-memory and disk-resident 
data---an essential requirement for scalability. We can leverage scalable dataframes such as \emph{Modin}~\cite{modin} that allows dataframes to exceed memory limitations. The extension would require integrating the \vta library in \emph{Modin} and adapting its column definition to include metadata schema. 
Another alternative is 
embedded analytical systems that tightly
integrate RDBMS with analytical tools and provide fast and efficient access to the data stored within them~\cite{raasveldt2020data}. \system employs PostgreSQL for session management and version control. Designing a storage layer that enables efficient data sharing between \vita sessions and the RDBMS is a possible research direction.}

\candidate{
\stitle{Efficient storage model.}
While \vitaframe currently supports data that fits in memory, in practice, the datasets will be larger. So, the storage layer of \system should enable operations on both in-memory and disk-resident 
data---an important requirement for scalability. Therefore, instead of Pandas, we can leverage scalable dataframes such as \emph{Modin}~\cite{modin} that supports both main-memory and persistent storage out-of-core, allowing intermediate dataframes to exceed memory limitations. The extension would require integrating the \vta library in \emph{Modin} and adapting its column definition to include metadata schema. Another alternative is 
embedded analytical systems that tightly
integrate RDBMS with analytical tools and provide fast and efficient access to the data stored within them~\cite{raasveldt2020data}. \system employs PostgreSQL for session management and version control. Designing a storage layer that enables efficient data sharing to allow for seamless passing of data back and forth between \vita sessions and the RDBMS is a possible research direction.}

\papertext{\stitle{Interaction at scale.}
\system should provide interactive responses even with larger datasets. One approach is to
draw samples of data progressively and then display approximate results. 
Approximate query processing techniques can support operations
like \code{aggregate}~\cite{babcock2003dynamic, agarwal2013blinkdb,acharya1999aqua} and \code{visualize}~\cite{rahman2017ve,hellerstein1997online}.
However, providing meaningful intermediate results progressively for operations like classification
or clustering is challenging---how to
determine model meta-parameters without
scanning the complete data? 
Progressive computation can be complemented by \emph{optimistic analytics}~\cite{moritz2017trust}, where precise computations run on the background as users explore approximate results. When there is a significant difference between
the approximate and precise results (\eg classification results vary from ground truth), the analyst can decide
which parts of the exploration have to be redone.
Larger datasets also impede direct data manipulation and necessitate the design of interactive and navigable representation of Table View~\cite{bendre2019faster}.}

\candidate{\stitle{Interaction at scale.}
One of the primary requirements of analytics tools like \system is to ensure interactivity even with larger datasets. One approach is to allow users to select a sample of data upfront via the \emph{selection} operator and then conduct the rest of the session on the sample for preliminary EDA---a common practice among data scientists~\cite{zhang2020teddy}. However, this approach still doesn't solve the scalability problem. 
An alternative is to pursue system driven sampling, where the goal is to 
draw samples of data progressively and then display approximate results. 
We call this approach Pro-\vita, \ie progressive \vita. 
How do we provide users with approximate, yet informative, intermediate responses? 
While we can leverage existing work from the DB community
related to approximate query processing to support operations
like \emph{aggregation}~\cite{hou1989processing,babcock2003dynamic, agarwal2013blinkdb,acharya1999aqua} and \emph{visualization}~\cite{rahman2017ve,fisher2012trust,hellerstein1997online}, it is not clear
how to provide meaningful intermediate results progressively for operations like classification
or clustering. For example, how to
determine crucial model meta-parameters without
having access to the complete data? 
Progressive computation can be complemented by \emph{optimistic analytics}~\cite{moritz2017trust} where precise computations run on the background 
as users explore approximate results. When there is a significant difference between the approximate and precise results (\eg classification results vary from ground truth), the analyst can decide which parts of the exploration have to be redone.
Larger datasets also impede direct data manipulation---a common problem of spreadsheet-style interfaces and necessitates design of interactive and navigable representation of the data~\cite{bendre2019faster}.}
% As users edit the data, for example, to relabel hundreds of instances, how can we retrain a model interactively?

\stitle{Versioning \vita sessions.}
The \system versioning system can maintain a version graph to keep track of fine-grained changes at the unit operation level. 
\system needs to consider the storage-latency trade-off as a user adds new nodes to the graph: storing entire data ensures faster session reconstruction at the cost of storage while storing delta between subsequent session reduces storage overhead at the cost of increased reconstruction time. 
Designing a fine-grained version control system for \system offers unique research challenges---besides \vitaframe, \system also needs to checkpoint
(a) the states of all the front-end components (\eg formatting like font, color, opacity of views), 
(b) the coordination mappings of views and composite selections, (c) user-defined operator pipelines and custom models, and (d) \vta commands in Notebook View. Existing systems address some of these challenges in isolation (\eg data~\cite{huang2017orpheusdb} and model~\cite{miao2016modelhub} versioning, workflow debugging~\cite{brachmannbyour,miao2016provdb}).

\stitle{\vta: coverage, accessibility, and automation.} 
Our goal is to increase \system's coverage of \vita workflows~\cite{liu2018bridging} by introducing new \vta operators and adding popular ML and NLP libraries~\cite{mlbazaar} as default operators in Operator View. We can further improve \system's extensibility by enabling users to add their custom-built models as new operators in Operator View and reuse later. To make \vta more accessible to a wider audience, we are working on integrating \vta with an interactive widget that allows users to issue \vta operators from Jupyter notebooks. Other goals include 
automatically generating \vita workflows given an analysis goal~\cite{bar2020automatically}, recommending next operator based on users' current workflow~\cite{yan2020auto}, and training
an autoregressive language model like \emph{GPT-3}~\cite{brown2020language} on \vta to automatically compose coordinated views or \system-style user interface based on user specifications in natural language.
%With increasing operator coverage, there will be newer challenges in \vta compiler design that ranges from mapping high-level specification to low level instructions and resolving ambiguities~\cite{satyanarayan2016vega}.
%\candidate{Such an interactive widget can enhance the notebook experience by allowing users to perform richer interaction with the data and computing resources. Other long term goals involve designing Auto-\vita systems that automatically generate \vita sessions given users' analysis goal~\cite{bar2020automatically} and developing recommendation systems that suggestion next operator based on users' current workflow~\cite{yan2020auto}. An ambitious goal is to train an autoregressive language model like \emph{GPT-3}~\cite{brown2020language} on \vta to automatically compose coordinated views or even a \system-style user interface based on user specifications in natural language.}

\stitle{\vita workflow optimization.}
Operator pipelines in a \vita workflow be executed in different order. For example, text reviews can be tokenized first and then cleaned or the other way around. However, tokenizing a cleaned text is less expensive due to it's smaller cardinality than the original text. We can leverage \vta to design a \vita workflow optimizer, similar to a query optimizer in databases. 
Other approaches for workflow optimization can use  
parallel execution similar to \emph{Modin}~\cite{modin},
enabling distributed processing of partitions of a \vitaframe and speed up computation.
\candidate{Designing an optimizer for progressive computation can be another interesting research direction~\cite{hou1988statistical,haas1996selectivity}.}

\stitle{Evaluation of \system.} While we demonstrate the expressivity and on-demand workflow authoring capabilities of \vta with several usage examples, we did not report the performance of other \system components. Future iterations should include experiments that evaluate the storage model's efficiency in managing large datasets, storage overhead and responsiveness of the versioning system, performance of \vta query optimizer, and overall interactivity of \system. Moreover, user studies should be conducted to evaluate the usability of \system.

\section{Related Work}\label{sec:related}

We now summarize existing work on \vita, data management for data science, and domain-specific algebra design.

\stitle{Visual interactive text analytics.}
\vita systems employ visualization techniques---both basic 
(\eg scatterplot, line chart, treemap) and complex (\eg wordcloud, steam graph, flow graph)~\cite{liu2018bridging}---to numerous uses-cases like review exploration~\cite{zhang2020teddy}, sentiment analysis~\cite{kucher2018state}, text summarization~\cite{carenini2006interactive}. While these earlier works highlight the appeal of integrating interactive visualization with text analysis, they lack the generalizability of \system, where users can employ an expressive algebra to author different \vita use-cases within a single system.

\stitle{Data management for \vita.}
Prior work focuses on designing systems for scalable computation (\eg scalable dataframe and query optimization~\cite{modin}, caching and prefetching for visualization~\cite{taokyrix}), storage models for efficient data access~\cite{tiledb,raasveldt2020data}.
We discussed related work on versioning~\cite{huang2017orpheusdb,miao2016modelhub,brachmannbyour,miao2016provdb}, approximate query processing~\cite{babcock2003dynamic, agarwal2013blinkdb,acharya1999aqua,rahman2017ve,hellerstein1997online} in earlier sections. \system builds on the earlier work with specific focus on developing an efficient storage model, enabling scalable computation, and performing fine-grained version control.

\stitle{Data model and algebra.}
Our work takes inspiration from existing algebras that provide well-founded semantics for relational databases~\cite{codd}, dataframes~\cite{modin,lara}, and interactive visualizations~\cite{stolte2002polaris,satyanarayan2016vega,satyanarayan2015reactive}. Here we introduce a new grammar for visual text data analytics  and  interactive view coordination, building on earlier work. To the best of our knowledge, \vta  is the first algebra defined for \vita.

\section{Conclusion}\label{sec:conclusion}
This paper presents our vision for \system, an integrated system that supports \vita workflows end to end. \system is designed based on several design considerations that we derived by identifying existing challenges in developing \vita systems. \system enables users to perform interactive text analysis in-situ---direct data manipulation (Table View), REPL-style analytics (Notebook View), and coordinated visual data exploration (Visualization  View).
We introduce a novel algebra for visual text analysis, \vta, that provides a suite of operators to author any \vita workflow on-demand and enable different modes of coordination among views.
We present our current progress in developing \system's underlying data management system and outline several research directions related to 
\vta extensibility and coverage, and storage, computation, and versioning of data and \vita workflows. Addressing these challenges requires interdisciplinary research efforts from the DB, NLP, HCI, and Visualization communities.

% \section{Conclusion}\label{sec:conclusion}
% \sajreview{In this paper, we outline our vision for \system, a system that captures the end-to-end flow of \vita. Using \system, users can perform  iterative/nonlinear analysis in-situ---direct data manipulation (\emph{table view}), REPL-style analytics (\emph{notebook view}), and coordinated visual data exploration (\emph{summary view})---without having to move back and forth between various tools.
% We also identify  a number of  challenges related to data management for \vita, \ie 
% efficient storage and computation models for heterogeneous data management and analysis, 
% provenance for 
% workflow and data re-use/reproduciblity, and
% an expressive grammar for \vita workflow abstraction and optimization. 
% We present our current progress 
% in developing \system's underlying data management system 
% and outline the key challenges that such a system should address next. These challenges are not only of interest to the DB community, but also require interdisciplinary research efforts spanning natural language processing, visualization, and human computer interaction.}

\bibliographystyle{ACM-Reference-Format}
\bibliography{paper}

\end{document}